\documentclass[preprint,12pt]{aastex} 
\usepackage{graphicx}
\usepackage{amsmath}
\usepackage{verbatim}
\usepackage{lineno}
\usepackage{natbib} 

\pdfoutput=1 
 
\newcommand{\bA}{\mathbf{A}}
\newcommand{\bB}{\mathbf{B}}
\newcommand{\bH}{\mathbf{H}}
\newcommand{\bn}{\mathbf{n}}
\newcommand{\bu}{\mathbf{u}}

\newcommand{\bw}{\mathbf{w}}

\newcommand{\bK}{\mathbf{K}}
\newcommand{\bM}{\mathbf{M}}
\newcommand{\E}{\mathbb{E}}

\begin{document}
\title{Invasion speeds for structured populations in fluctuating environments}
\author{Sebastian J. Schreiber}
\author{Maureen E. Ryan}
\affil{Department of Evolution and Ecology and the Center for Population Biology\\ 
University of California, Davis, California 95616}

\bibliographystyle{plainnat}

\begin{abstract}
\baselineskip 20pt
\linenumbers
We live in a time where climate models predict future increases in environmental variability and biological invasions are becoming increasingly frequent. A key to  developing  effective responses to biological invasions in increasingly variable environments will be estimates of their rates of spatial spread and the associated uncertainty of these estimates. Using stochastic, stage-structured, integro-difference equation models, we show analytically that invasion speeds are asymptotically normally distributed with a variance that decreases in time. We apply our methods to a simple juvenile-adult model with stochastic variation in reproduction and an illustrative example with published data for the perennial herb, \emph{Calathea ovandensis}. These examples buttressed by additional analysis reveal that increased variability in vital rates simultaneously slow down invasions yet generate greater uncertainty about rates of spatial spread. Moreover, while temporal autocorrelations in vital rates inflate variability in invasion speeds, the effect of these autocorrelations on the average invasion speed can be positive or negative depending on life history traits and how well vital rates ``remember'' the past. 
\end{abstract}

\maketitle

\vskip 0.2in 
\centerline{\textbf{Word count}: 4,649 without Appendix; 4,807 with Appendix}
\centerline{\textbf{Figures}: 4 \textbf{Tables}: 0 \textbf{References}: 50}
\newpage
\baselineskip 20pt
\linenumbers
\section*{Introduction}

Anthropogenic forces are changing the temporal distribution of environmental fluctuations and accelerating the rate at which species are being introduced into non-native habitat. General circulation models predict increased variability in temperatures and precipitation and these changes are likely to increase variability in vital rates of many species~\citep{easterling-etal-00,tebaldi-etal-06,boyce-etal-06}. Alternatively, human activities such as agriculture, recreation, and transportation are spreading species beyond their natural dispersal barriers~\citep{elton-58,kolar-lodge-01}. While most of these accidental or intentional introductions fail, the successful invaders can have devastating impacts on human health and native biodiversity~\citep{kolar-lodge-01}.  To manage these impacts, it is essential to understand the rate of range expansion--the invasion speed--of these invaders. Here, we bring together the theory of stochastic demography and invasion speeds to provide a general framework to estimate invasion speeds for structured populations in a variable environment. 

Stochastic demography is concerned with understanding population growth and distribution when vital rates vary in time due to stochastic fluctuations in environmental variables~\citep{boyce-etal-06}. In their simplest guise, models of stochastic demography are the form $\bn_{t+1}=\bA_t \bn_t$ where $\bn_t$ is the vector of population abundances in the different stages at time $t$ and $\bA_t$ is a projection matrix whose entries describe fluxes between stages due to combinations of growth, survivorship, and reproduction. These models play a critical role in identifying to what extent variation in vital rates alter the stochastic growth rate of a population~\citep{tuljapurkar-90,tuljapurkar-etal-03,haridas-tuljapurkar-05,morris-etal-08}.  For example, \citet{morris-etal-08} analyzed multiyear demographic data for 36 plant and animal species  and found that the stochastic growth rate of short-lived species (insects, annual plants, and algae) are more negatively affected by increased variation in vital rates than  longer-lived species (perennial plants, birds, ungulates). Correlations between vital rates within a year or autocorrelations in vital rates in successive years can significantly effect stochastic growth rates. Analytic approximations imply that negative correlations between vital rates can buffer the effects of demographic stochasticity and thereby increase stochastic growth rates \citep{tuljapurkar-90,boyce-etal-06}. In contrast, temporal autocorrelations  can increase or decrease stochastic growth rates, and increases in these autocorrelations can have larger impacts on growth than increases in inter-annual variability of vital rates~\citep{tuljapurkar-90,tuljapurkar-haridas-06}. 

For populations not experiencing an Allee effect, a positive growth rate at low densities is necessary for establishment and range expansion. An important class of models for predicting rates of range expansion are integro-difference equation (IDE) models~\citep{kot-etal-96} in which a dispersal kernel describes the likelihoods that individuals move between locations. For diffusive movement, this kernel is a Gaussian and the rate of spread equals $2\sqrt{r D}$ where $D$ is the diffusion rate and $r$ is the intrinsic rate of growth of the population~\citep{kot-etal-96}; the same rate of spread first derived by \citet{fisher-37} for reaction-diffusion equations. While this estimate of rate of spread has been applied to many species~\citep{hengeveld-94}, field estimates of dispersal kernels typically are leptokurtic  and not Gaussian~\citep{kot-etal-96} and, consequently, this earlier work may underestimate invasion speeds. Using stochastic counterparts to these IDE models, \citet{neubert-etal-00} showed that serially uncorrelated, stochastic fluctuations generate normally distributed invasion speeds whose variance decays to zero. Hence, invasion speeds may exhibit  unpredictable transients in fluctuating environments. On the other hand,  \citet{neubert-caswell-00} developed methods to estimate invasion speeds for stage-structured IDE models that have been invaluable for identifying how stage-specific vital rates constrain rates of spatial spread~\citep{caswell-etal-03,jacquemyn-etal-05, jongejans-etal-08}. However, the dual effects of demographic and temporal heterogeneity on rates of range expansion remains to be understood. 


Here, we provide a framework for analyzing the simultaneous effects of environmental fluctuations and demographic structure on invasion speeds. We begin by reviewing the work of \citet{neubert-caswell-00} in constant environments. We extend these models to allow for temporal variation in the projection matrices and dispersal kernels and provide a formula for asymptotic invasion speeds and normal approximations that describe the variation in these invasion speeds over finite time horizons. Applying these results to two examples, we illustrate how our results allow one to address questions like ``how does the magnitude of variability influence asymptotic invasion speeds?'' and ``how do temporal autocorrelations influence the uncertainty in predicting invasion speeds?''   

\section*{Constant environments}

\citet{neubert-caswell-00} analyzed invasion speeds for IDE models for stage-structured populations in constant environments. These models consider  structured populations  living on a continuous one-dimensional habitat and consisting of $m$ stages where $n^i_t(x)$ is the density of stage $i$ at time $t$ in location $x$. Let $b^{ij}(n^1(x),\dots,n^m(x))$ be the contribution of stage $j$ individuals to stage $i$ individuals at location $x$. Let $k^{ij}(x)$ be the probability density function for the displacement $x$ moved by an individual transitioning from stage $j$ to $i$. Under these assumptions, the dynamics of the population are given by 
\[
n^i(x)= \int_{-\infty}^\infty \sum_{j=1}^m k^{ij}(x-y)b^{ij}(n^1_t(y),\dots,n^m_t(y)) n_t^j (y)\,dy
\]
To simplify the notation, let $\bn_t(x)=(n^1_t(x),\dots,n^m_t(x))'$ where $'$ denotes transpose be the vector of population abundances at time $t$ and location $x$. Let $\bB(\bn_t(x))$ and $\bK(x)$ denote the $m\times m$ matrices with entries $b^{ij}(\bn_t(x))$ and $k^{ij}(x)$, respectively. With this notation, we get the simplified equation
\begin{equation}\label{eq:one}
\bn_{t+1}(x)=\int_{-\infty}^\infty \left[ \bK(y-x)\circ \bB(\bn_t(y))\right] \bn_t (y)\,dy
\end{equation}
where $\circ$ denotes the Hadamard product i.e. component wise multiplication. 

When the population is unstructured (i.e. $m=1$), model \eqref{eq:one} has 
traveling waves-solutions	that maintain a fixed shape in space and move at a constant speed. If the growth function $b(n)n$ increases with
population density,  $b(n)$ is decreasing (i.e. no Allee effect), and the dispersal kernel possesses a moment-generating function $m(s)=\int_{-\infty}^\infty k(x)e^{-sx}\,dx$ for $0\le s< \hat s$, 
then the traveling wave has an asymptotic speed
\[
c^*=\min_{0<s<\hat s}\frac{1}{s}\ln \left[ b(0) m(s)\right]
\]
A population initially concentrated in a finite region of space will never spread faster than $c^*$ and asymptotically will spread at a rate of exactly $c^*$ ~\citep{weinberger-82}. The linearization conjecture states that the speed of invasion
for a nonlinear model is governed by its linearization at low population densities, as long as there are no Allee effects and no long-distance density dependence.
This conjecture is supported by many numerical studies. 

Relying on the linearization conjecture, \citet{neubert-caswell-00} derived a formula for traveling wave speeds for structured populations. This derivation makes four assumptions: (1) the matrices $\bB(\bn)$ are non-negative and primitive, (2) $\bA=\bB(0)$ has a dominant eigenvalue $\rho(\bA)$ that is greater than one, (3) there is negative density dependence i.e. $\bB(\bn)\le \bA\bn$ for all $\bn\ge 0$ where inequalities are taken componentwise, and (4) the kernels $k^{ij}(x)$ have  moment generating functions $m^{ij}(s)$ defined on some maximal interval $0\le s<\hat s$. Under these assumptions, \citet{neubert-caswell-00} showed  the asymptotic wave speed is given by 
\begin{equation}
c^* = \min_{0<s<\hat s} \frac{1}{s} \log \rho (\bA\circ \bM(s))
\end{equation}
where $\bM(s)$ is the matrix of moment generating functions $m^{ij}(s)$. 

\section*{Fluctuating environments}

To account for temporal variation in environmental conditions and dispersal rates, we allow $\bB_t(\bn)$, and $\bK_t(x)$ to depend on time. In which case,
\begin{equation}\label{eq:two}
\bn_{t+1}(x)=\int_{-\infty}^\infty \left[ \bK_t(x-y)\circ \bB_t(\bn_t(y))\right] \bn_t (y)\,dy
\end{equation}
In order to make use of the linearization conjecture, we place four assumptions on \eqref{eq:two}. First,  the populations exhibit negative density dependence in which case $\bA_t:=\bB_t(0)\le \bB_t(\bn)$ for all $\bn\ge 0$.  Second, there exists $n_0>0$ such that $\bA_{n_0}\dots \bA_1$ has all positive entries with probability one. This assumption ensures the effects of the initial distribution of the population disappear in the limit (demographic weak ergodicity). Third, there exists a  $\hat s>0$ such that the kernels $k^{ij}_t(x)$ have moment generating functions $m^{ij}_t(s)$ defined for $0\le s < \hat s$. We always choose $\hat s$ to be the maximum of these values. Without this assumption, solutions may accelerate and have no well-defined asymptotic wave speed~\citep{kot-etal-96}. To state the fourth assumption, let $\bM_t(s)$  be the matrix with entries $m^{ij}_t(s)$ and $\bH_t(s)=\bM_t(s)\circ \bA_t$. We assume that the sequence of matrices, $\bH_0(s),\bH_1(s),\dots$  are stationary and ergodic and  $\E[\max\{\ln\|\bH_t(s)\|,0\}]<\infty$ for all $0\le s<\widehat s$. This assumption encompasses many models of environmental fluctuations including periodic, quasi-periodic, irreducible Markovian, and auto-regressive models. Under this assumption, the random version of the Perron-Frobenius theorem \citep{arnold-etal-94} implies that  for any $\bn>0$ and $\bw>0$ (i.e. all components are non-negative and at least one is positive)
\begin{equation}\label{eq:gamma}
\lim_{t\to\infty} \frac{1}{t} \ln \langle\bH_{t-1}(s)\dots \bH_0(s) \bn, \bw\rangle =\gamma(s) \mbox{ with probability one}
\end{equation}
where $\gamma(s)$ is the dominant Lyapunov exponent associated with this random product of matrices and $\langle \cdot, \cdot \rangle$ denotes the standard Euclidean inner product. When $s=0$ and $\bw=(1,\dots,1)'$, this Lyapunov exponent
\[
\gamma(0)=\lim_{t\to\infty} \frac{1}{t} \ln \langle \bA_{t-1}(s)\dots \bA_0(s) \bn,\bw \rangle
\]
describes the growth rate of the total population size when rare \citep{tuljapurkar-90}. Since an invasion can only proceed if the population has the capacity to exhibit growth, our final assumption is that $\gamma(0)>0$. 

\begin{figure}[t!!]
\begin{center}
\includegraphics[width=5in]{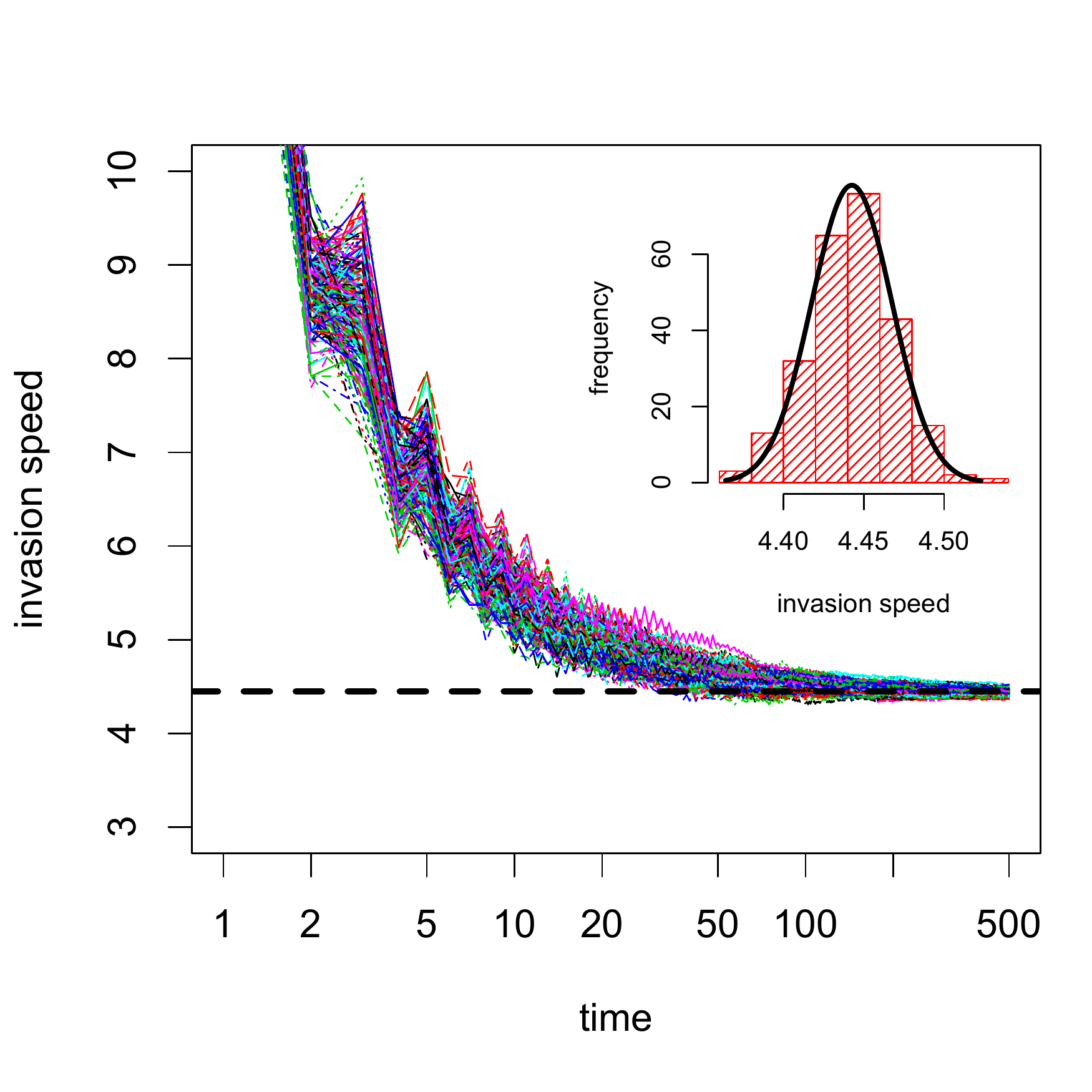}
\caption{The temporal dynamics of the wave speed $\frac{X_t-x_0}{t}$ for $250$ simulations of the non-linear juvenile-adult model. The front of the wave was determined by a threshold of $n_c=0.001$ with equal weight on both stages i.e. $\bw=(1,1)'$. The dashed line is the predicted asymptotic wave speed in \eqref{eq:speed}. In the inset, a histogram of the waves speeds at $t=500$ with the predicted normal approximation from the linearization. Parameter values are $\rho=0$, $\mu=\ln 40$, $\sigma=0.5$, $a=1$, $s_J=0.3$, and $s_A=0.4$. }\label{fig:one}
\end{center}
\end{figure}

\begin{figure}[t!!]
\begin{center}
\includegraphics[width=6.5in]{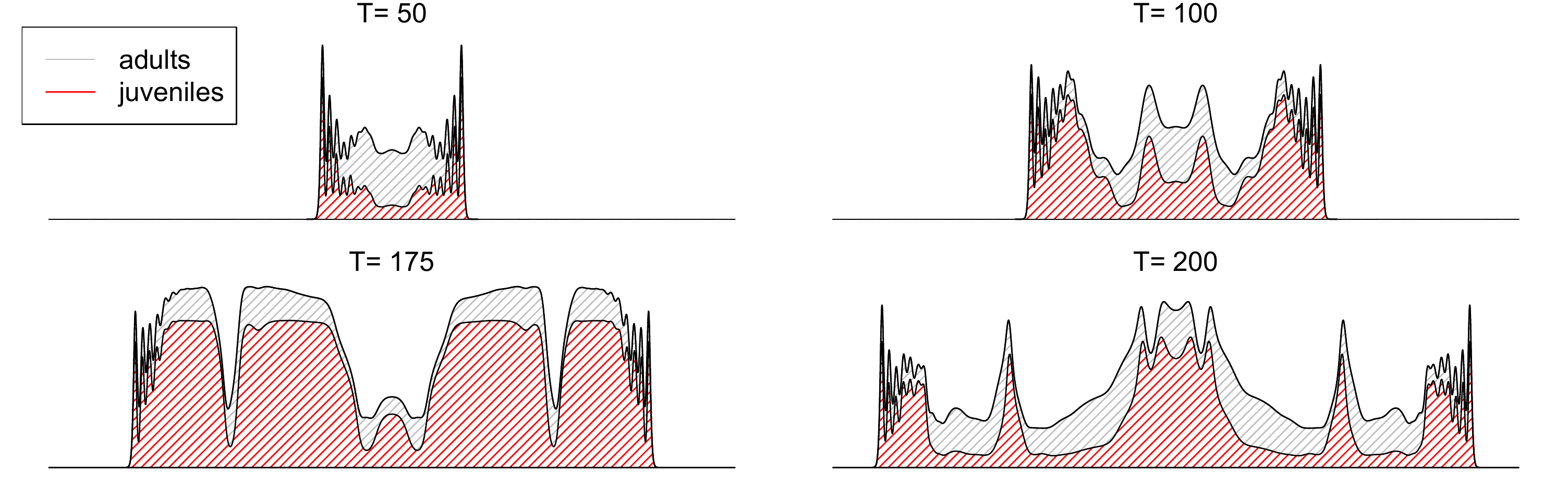}
\caption{Spatio-temporal dynamics of range expansion for the juvenile-adult model. Spatial distribution and abundance of juveniles (in shaded red) and adults (in shaded grey) plotted at the indicated times. Parameters ($\rho=0$, $\mu=\ln 40$, $\sigma=0.1$, $a=1$, $s_J=0.3$, and $s_A=0.4$) are such that local dynamics are chaotic. }\label{fig:two}
\end{center}
\end{figure}

When demographic or dispersal rates vary in time, the rate of spatial spread will vary in time. To quantify this rate of spread, assume there is a weighting $\bw>0$ of the different stages and a weighted population abundance, $n_c>0$, above which the population is observable. At time $t$, let $X_t$ denote the location furthest from the invasion's origin at which $\langle \bn_t(s), \bw \rangle \ge n_c$. The average speed $C_t$ of the invasion by time $t$ is given by $\frac{X_t-x_0}{t}$ where $x_0$ is the introduction site. In  Appendix A we show that with probability one,  $\frac{X_t-x_0}{t}$ is asymptotically bounded above by  
\begin{equation}\label{eq:speed}
c^*=\min_{0\le s<\hat s} \frac{\gamma(s)}{s}
\end{equation}
whenever $\bn_0(x)$ has compact support. The linearization conjecture and our numerical results (Fig.~\ref{fig:one}) suggest that $c^*$ is not only an upper bound but in fact equals the asymptotic wave speed with probability one. In the special case of periodic environmental fluctuations (i.e.  there exists a natural number $p$ such that $\bA_{t+p}=\bA_t$ and $\bK_{t+p}(x)=\bK_t(x)$ for all $t$ and $x$),  the asymptotic invasion speed is given by
\[
 c^*= \min_{s>0}\frac{1}{sp}\ln \rho\left(\prod_{t=1}^p \bA_t \bM_t(s)\right)
\]
where  $\rho$ denotes the dominant eigenvalue of a matrix. 

Under the additional assumption that matrices $\bH_t(s)$ are rapidly mixing (see, e.g., \citet{heyde-cohen-85}), Appendix B shows that there exists $\sigma>0$ such that the average speeds $\frac{X_t-x_0}{t}$ are asymptotically normal with mean $c^*$ and standard deviation $\sigma/(s^*\sqrt{t})$. This approximation is consistent with numerical simulations of the nonlinear models (inset of Fig.~\ref{fig:one}).  In the special case of an unstructured population $m=1$, this result extends \citet{neubert-etal-00}'s work on uncorrelated environments to correlated environments. In this case, the central limit theorem for stationary sequences of random variables \citep{durrett-96} implies
\begin{equation}\label{eq:cov}
\sigma^2= \mbox{Var}[\ln\bH_1(s^*)]+2 \sum_{i=2}^\infty \mbox{Cov}[\ln\bH_1(s^*),\ln\bH_i(s^*)]
\end{equation}
where $s^*$ is such that $c^*=\gamma(s^*)/s^*$. For unstructured populations, equation \eqref{eq:cov} implies that positive temporal autocorrelations increase the variability in asymptotic wave speeds while negative autocorrelations reduce this variability. In contrast, as $\gamma(s)=\E[\ln H_1(s)]$,  temporal autocorrelations have no effect on the mean $c^*$ invasion speed for unstructured populations.  

To illustrate the applicability of our results to structured populations, we consider two examples. The first example is a simple juvenile-adult structure model with continuous variation in fecundity rates. We use this example to illustrate the accuracy of the linear approximation, the dynamics of spatial spread, and the effects of noise amplitude and color on invasion speeds and stochastic growth. In the second example, we analyze  the effects of environmental stochasticity on invasion speeds for an herbaceous perennial herb, \emph{Calathea ovandensis} (\emph{Marantaceae}). This example was studied for constant environments by \citet{neubert-caswell-00}. Here, we illustrate the effects of the frequency of poor environmental conditions and temporal correlations on invasion speeds and their uncertainty using empirical data. 

\begin{figure}[t!!]
\begin{center}
\begin{tabular}{cc}
\includegraphics[width=3.25in]{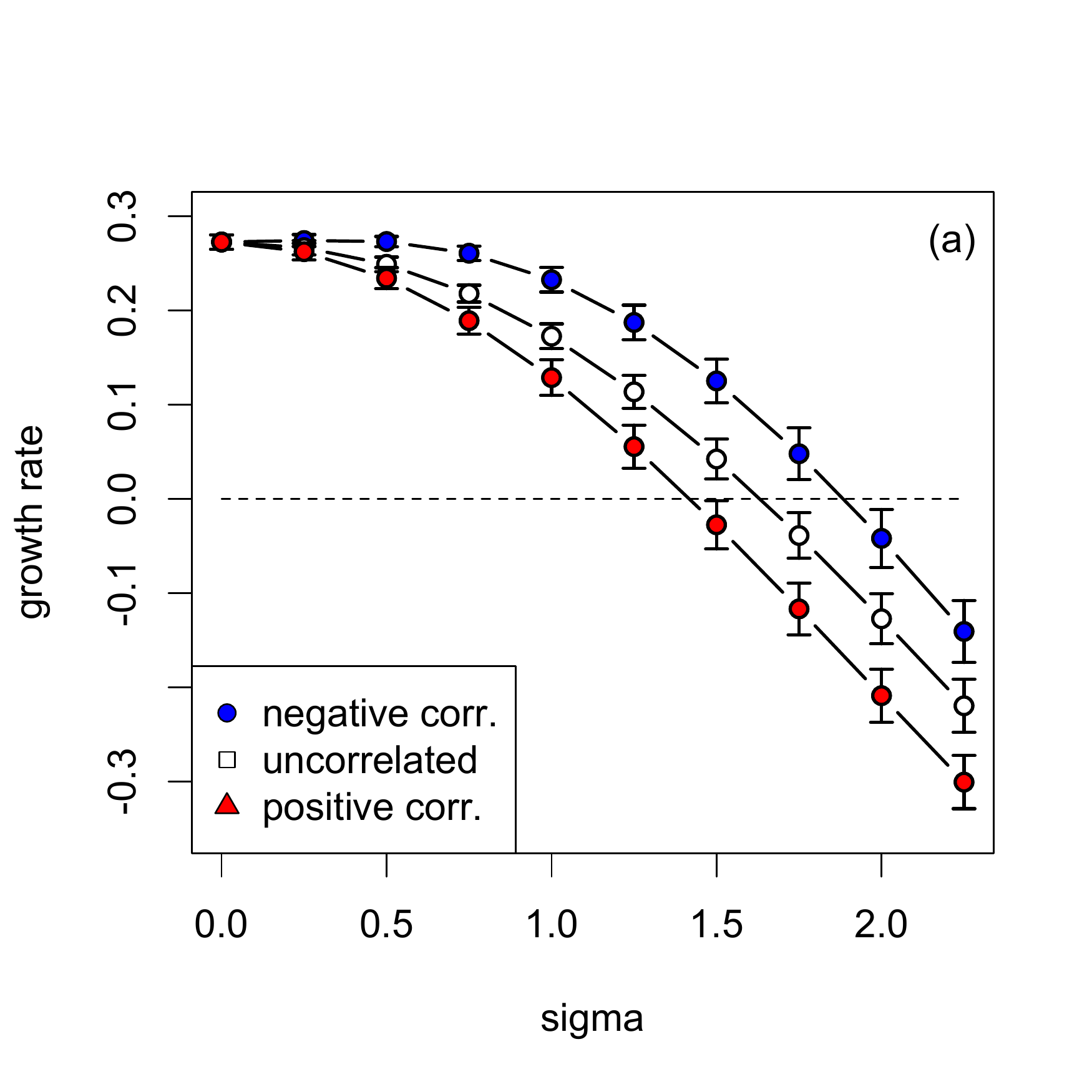}&\includegraphics[width=3.25in]{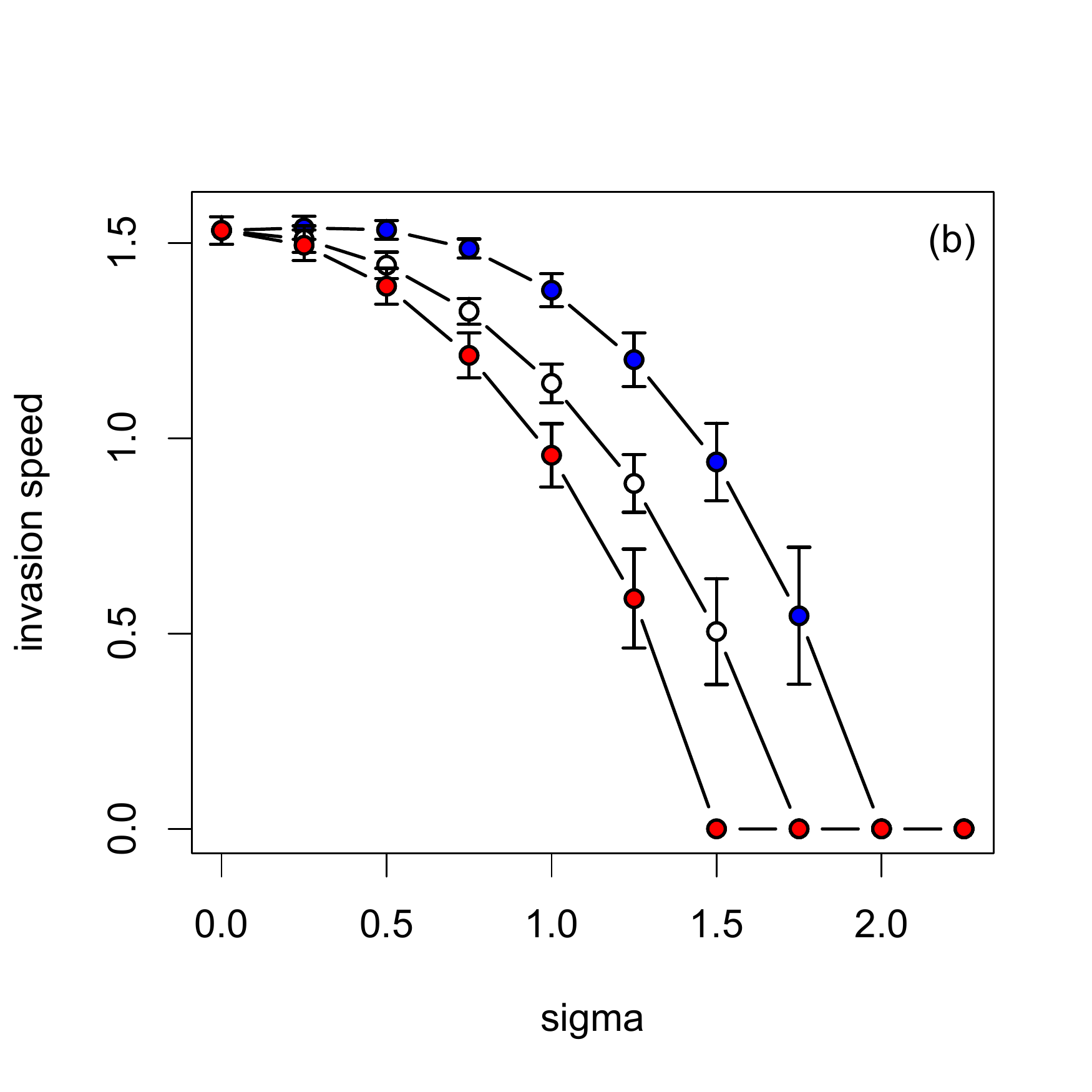}
\end{tabular}
\end{center}
\caption{The stochastic growth rate $\gamma$ and invasion speed $c^*$  plotted as a function of the standard deviation of the log fecundity $\log f_t$. Length of error bars correspond to the standard deviation in the growth rate after $400$ generations. The  blue, white, and red curves correspond to temporal correlations of $\rho=-0.5, 0$ and $0.5$. Parameters are such that  $\E[f_t]=4$, $s_J=0.3$, and $s_A=0.4$. Simulations for estimates ran for 10,000 generations.}\label{fig:three}
\end{figure}

\paragraph*{Example: Fluctuating fecundities.}
We consider a population with two stages, juvenile and adult, in which dispersal occurs following reproduction. We assume that the per-capita fecundity $f_t \exp(-a\,n^2_t)$ is density-dependent where $n_t^2$ is the adult density at time $t$, $f_t$ is the maximal fecundity at time $t$, and $a>0$ measures the intensity of intraspecific competition. Let  $s_J$ denote the fraction of juveniles surviving to adulthood, and $s_A$  denote the fraction of adults that survive to the next year. Under these assumptions, 
\[
\bB_t(\bn)=\begin{pmatrix}0&f_t \exp(-a\,n^2)\\ s_J& s_A\end{pmatrix}
\]
We assume that $\log f_t$ are normally distributed with mean $\mu$, variance $\sigma^2$, and temporal correlation $\rho$ between $\log f_t$ and $\log f_{t+1}$. Dispersal follows birth with a Laplacian distribution with variance $2b^2$. Hence, the moment generating matrix for this model is 
\[
M(s)=\begin{pmatrix}1&\frac{1}{1-(bs)^2} \\ 1& 1\end{pmatrix}
\]
which is defined for $0\le s \le \hat s = 1/b$. When adults are highly fecund, the interaction between locally unstable population dynamics and environmental stochasticity can generate complex spatial-temporal patterns of range expansion (Fig.~\ref{fig:two}). None the less, simulations of the full nonlinear mode suggest that linearization conjecture holds (Fig.~\ref{fig:one}). 

To understand the effects of temporal variation and correlations on wave speeds and population growth, we computed $\gamma(0)$, $c^*$, and standard deviations of both of these quantities for a range of $\mu$ and $\sigma^2$ values. In all of these simulations, we held the mean fecundity $\E[f_t]=\exp(\mu-\sigma^2/2)$ constant at $4$. The simulations show that the population growth rate $\gamma(0)$ and the asymptotic wave speed decreases both with increasing temporal variation and increasing temporal autocorrelations (Fig.~\ref{fig:three}). In particular, when this variation is too high, the population has a negative stochastic growth rate and does not propagate. Autocorrelations have little effect on the variability of the stochastic growth rates, but variation in invasion speeds increases with  autocorrelations. 

\paragraph*{Example:\emph{ Calathea ovandensis}.}

 \emph{Calathea ovandensis} is an understory monocot, found in neotropical lowland rain forests~\citep{horvitz-schemske-95}. Deciduous during the dry season, \emph{C. ovandensis} plants reinitiate growth from rhizomes during the rainy season and bear fruit capsules, which dehisce to expose large seeds that fall to the forest floor~\citep{horvitz-beattie-80}. Seeds are myrmecochorous (adapted for ant dispersal), bearing oily white arils (eliaosomes) that are considered arthropod prey mimics (Carroll and Janzen 1973).  Ponerine ants carry seeds to their nests like prey, remove and consume the arils, and bury seeds in nitrogen-rich Ògarbage pilesÓ of dead insect prey near the nest entrance ~\citep{horvitz-beattie-80,horvitz-schemske-95}. 

At Horvitz and Schemske's study site in southern Mexico, four ant species act as disperal agents: \emph{Pachycondyla apicalis},\emph{ P. harpax}, \emph{ Solenopsis geminata}, and \emph{Wasmannia auropuncata} ~\citep{horvitz-schemske-95}. \emph{Calathea ovandensis} do not propagate vegetatively, and empirical distributions of seedlings are well matched to ant dispersal distances \citep{horvitz-schemske-86}. ~\citet{horvitz-schemske-95} measured demographic rates, constructing 16 projection matrices for four sites over four years, and found that \emph{C. ovandensis}' highest population growth rates occurred during el Ni\~{n}o years and in plots affected by treefall gaps ~\citep{horvitz-schemske-95,horvitz-etal-97}.
 
We selected 3 stage-structured matrices that represent 1) the highest population growth, observed during an el Ni\~{n}o year ($\lambda=1.2477$; plot 2, 1982-83; also used in \citet{neubert-caswell-00}), 2) lowest (negative) growth in the same plot ($\lambda=0.9051$; plot 2; 1984-85), and 3) the worst year-plot combination overall ($\lambda=0.7356$; plot 3, 1984-85). Individuals are classified as seeds, seedlings, juveniles, pre-reproductives, and small, medium, large and extra-large adult plants (for life cycle graph, see \citet{neubert-caswell-00}). Seed dispersal involves transitions from most stages. We follow \citet{neubert-caswell-00} in assuming a Laplace distribution of dispersal distances of four ant species and weighting dispersal by the relative abundance of each species ~\citep{horvitz-schemske-95,neubert-caswell-00}. This treatment assumes that seed fates are independent of dispersal agent \citep{neubert-caswell-00}, and yields the dispersal kernel matrix
\[
K(x-y)=\sum_{i=1}^4 p_i K_i(x-y)
\]
where $p_i$ are the proportion of seeds dispersed by each ant species, and $K_i(x-y)$ is a matrix whose first row contains Laplace dispersal kernels with generating functions $m_i(s)=1/(1-b_i^2s^2)$ with parameter $b_i$, representing the mean distance a seed is moved by a particular ant species as reported by \citet{horvitz-schemske-86}, and Delta functions for the remaining entries \citep{neubert-caswell-00}. The moment-generating function matrix is
\[
M(s)=\sum_{i=1}^4 p_i M_i(s)
\]
where $M_i(s)$ is the moment generating function for $K_i(x)$. 

\begin{figure}[t!!]
\hskip-0.5in\includegraphics[width=8in]{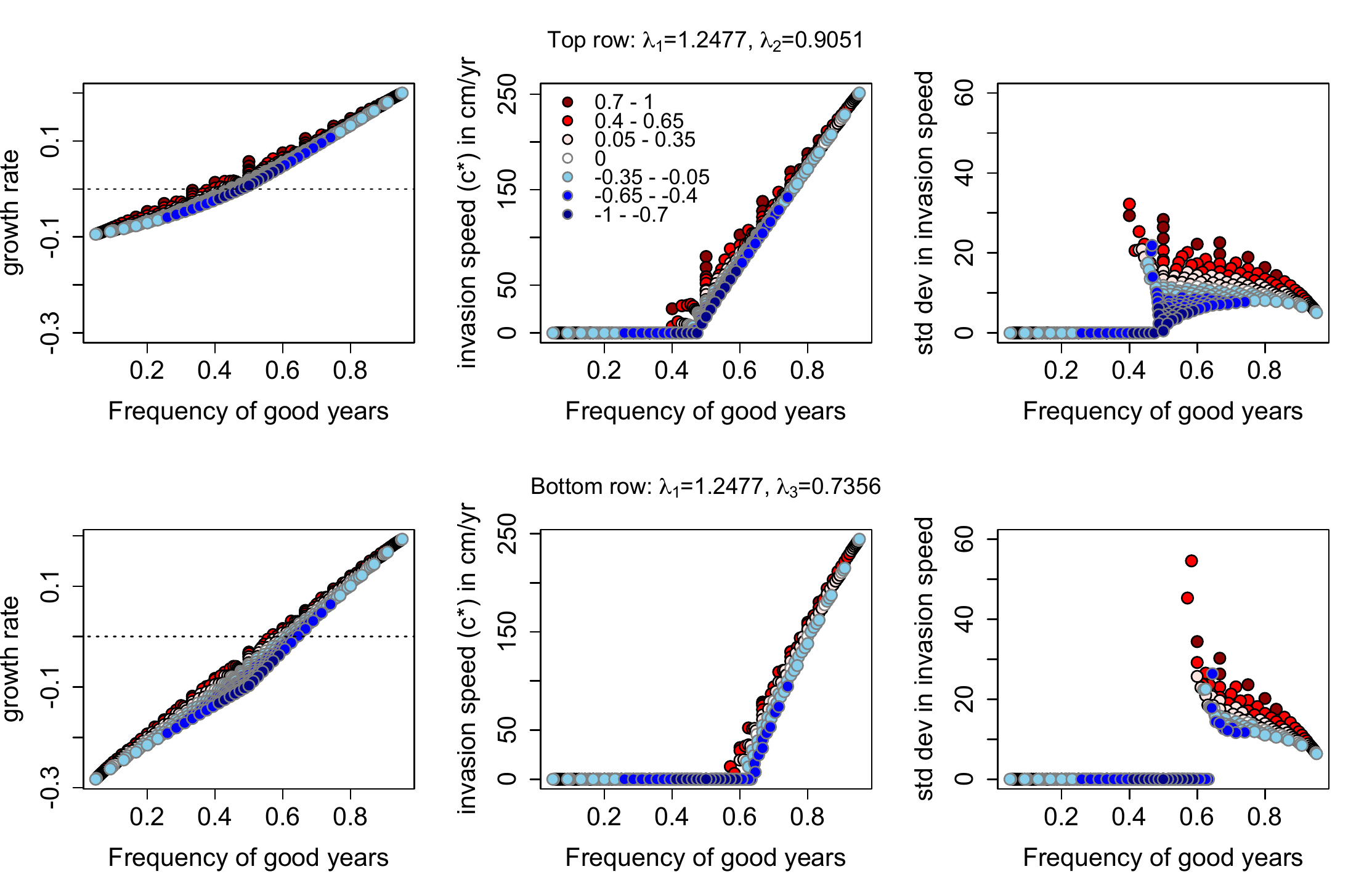}
\caption{ Stochastic growth rates and invasion speeds ($c^*$) for \emph{Calathea ovandensis}. Stochastic growth rates (left column), asymptotic invasion speeds (middle column), and standard deviation in invasion speeds after 400 years (right column) are plotted as a function of the frequency $q/(q+p)$ of good years for all probabilities $p$ and $q$ of switching from ``good'' to ``bad'' years and vice versa. Temporal correlations associated with the different $p,q$ combinations are plotted in color as indicated in the legend. Figures in the top row correspond to random fluctuations between environmental states corresponding to the highest population growth and the lowest population growth in the same plot (i.e. $\lambda = 1.2477$ and $\lambda = 0.9051$). Figures in the bottom row correspond to random fluctuations between environmental states corresponding to the highest and lowest overall population growth across all sites (i.e. $\lambda=1.2477$ and $\lambda=0.7356$). Simulations ran for 100,000 generations.}\label{fig:ants}
\end{figure}
	
We ran two types of simulations representing (1) random fluctuations between environmental states corresponding to the highest population growth and the lowest population growth in the same plot (i.e. randomly varying between the matrices with dominant eigenvalues $\lambda=1.2477$ and $\lambda= 0.9051$) and (2) random fluctuations between environmental states corresponding to the highest and lowest overall population growth across all sites (i.e. $\lambda=1.2477$ and $\lambda=0.7356$). For each of these simulations, switches between ``good'' and ``bad'' years were modeled by a Markov chain in which the probability of transition from a bad year to a good year is $q$ and the probability of transition from a good year to a bad year is $p$. For this model of the environmental dynamics, the long-term frequency of good years is $\frac{q}{q+p}$ and the temporal autocorrelation between environmental states is $1-q-p$. For both scenarios, simulations were run for all combinations of $p$ and $q$ values to determine the combined effects of frequency of good years and temporal correlations on stochastic growth and invasion speed.

Both the stochastic growth rate and invasion speed were strongly correlated with the overall frequency of good years, while correlations between environmental states had a relatively small effect on asymptotic growth rates and invasion speeds (Fig.~\ref{fig:ants}). However, positive correlations between environmental states did increase growth rates and invasion speeds, while negative correlations reduced both (left and middle column of Fig.~\ref{fig:ants}). For growth rates, the effect of correlations was greatest when the frequency of good and bad years were equal ($q/(q+p)=0.5$) (left column  of Fig.~\ref{fig:ants}), presumably since the largest number of combinations of $(p,q)$ values can generate this frequency. Environmental correlations had the greatest impact on invasion speeds near the threshold between positive and zero growth (middle column of Fig.~\ref{fig:ants}), where invasion speeds also exhibited the greatest variance (right column  of Fig.~\ref{fig:ants}). As a result, near the threshold in parameter values between positive and negative growth rates (and hence positive or zero invasion speed), the magnitude and sign of correlations between environmental states determined whether populations grew and spread or declined. For example, for scenario one (top row  of Fig.~\ref{fig:ants}), strong positive correlations allowed the population to grow when only 35\% of years were good, while strong negative correlations generated negative growth rates under the same scenario even at nearly even proportions of good and bad years. 

Like growth rates and invasion speed, variance in invasion speed was also strongly affected by both environmental correlation and the frequency of good years (right column of Fig.~\ref{fig:ants}). Most notably, invasion speed variance increased approaching the threshold of zero population growth rates in both scenarios, though most dramatically in the second (bottom row of Fig.~\ref{fig:ants}), and declined as the proportion of good years increased. Variance in invasion speed also increased as environmental correlations shifted from strongly negative to strongly positive. Hence, the combinations of environmental states that yielded the fastest invasion speeds also showed the highest variance in invasion speed. 

Comparing between scenarios, the scenario with a higher average growth rate between years (top row  of Fig.~\ref{fig:ants}) exhibited a lower threshold for positive growth rates with respect to the frequency of good years and overall lower variance in invasion speeds, as expected. Since positive invasion speeds increased roughly linearly with the proportion of good years, a necessary outcome was that each increment of change in the proportion of good years led to a greater proportional increase in invasion speed in the ``worse'' scenario (bottom row  of Fig.~\ref{fig:ants}) than the scenario in which average growth rate across years was higher (top row  of Fig.~\ref{fig:ants}). In the context of real systems, this means that we might expect to observe particularly steep increases in invasion speed in response to increased frequency of good years in regions where bad years are particularly bad.  In particular, changes in the frequency of El Ni\~{n}o years or in the temporal correlation between El Ni\~{n}o events, could substantially affect rates of spread of \emph{Calathea}.

\section*{Discussion}

How environmental variation influences population dynamics and biodiversity is a central question in ecology \citep{andrewartha-birch-54,may-75, tuljapurkar-82,chesson-00,holyoak-etal-05}. In our era of increasing climate variability and accelerating global transport of species at rates and distances far beyond their innate capacity~\citep{vitousek-etal-97}, this question is not only academic but also directly relevant to wildlife management and the maintenance of regional biodiversity. The ability to forecast and mitigate impacts of successful invaders, and changes in the distribution of native species, relies on an enhanced understanding of the potential impacts of environmental variability on population growth rates and invasion speeds~\citep{neubert-etal-00, collingham-huntley-00}. By extending a class of stage-structured integrodifference equation models to variable environments, we developed and applied a methodology for examining how environmental variability and temporal correlations impact stochastic growth rates, invasion speeds, and uncertainty in invasion speed predictions.

For structured populations in fluctuating environments, our analysis reveals that invasion speeds are approximately normally distributed and identifies data requirements for estimating the mean and variance of invasion speeds. As shown by \citet{neubert-etal-00} for unstructured populations in serially uncorrelated environments, we found that variance in invasion speeds decay in time. Hence, the greatest uncertainty in predicting the spatial extent of an invader occurs in the earliest stages of spatial spread.   The data requirements for quantifying this uncertainty are two-fold. To capture the effects of environmental fluctuations on demography, estimates of survivorship, fecundity, and transition rates between different stages are needed for the range of environmental conditions  most likely to be encountered by the population~\citep{caswell-01,boyce-etal-06}. Our methods also require estimates for the  dispersal kernel associated with each demographic transition under an appropriate range of environmental conditions. While getting such estimates  is challenging, our methods can help focus empirical efforts by identifying which stage-specific demographic rates have the largest effect on the mean and variance of the invasion speed. More generally, they can be used as an exploratory tool to identify key features of how environmental variability and stage-structure interact to determine invasion speeds.

Our examples illustrate several important general effects of environmental variability on invasion speeds, and point to the ways in which the outcomes of environmental-demographic interactions might vary between species. The results of our two-stage example with fluctuating fecundities showed growth rates and invasion speeds declining as the magnitude of variance in vital rates increased. This result is consistent with earlier work showing that increases in the magnitude of environmental (e.g. climate) variability tend to reduce population growth rates~\citep{lewontin-cohen-69}. Theory and empirical findings predict that the magnitude of these effects on population growth will depend on a species' life history: short-lived species are more negatively affected by increased variation in vital rates than longer-lived species~\citep{chesson-85,morris-etal-08}. Although not investigated here, it seems reasonable to conjecture that these life history traits will have similar effects on invasion speeds. 

While increasing environmental variability reduces growth and invasion speed, correlations between environmental states can have a substantial impact on the direction and variance in growth and wave speed. Our analysis of a general unstructured model and two structured models shows that positive temporal correlations between environmental states (reddened spectrum in figures) created greater variability in growth rates and invasion speeds, even at low levels of environmental noise. This finding is consistent previous work, which shows potentially strong effects of positive temporal correlations in vital rates on variance in population growth~\citep{tuljapurkar-orzack-80, tuljapurkar-82, runge-moen-98}. In contrast to the consistent effect of correlations on variance, the direction of effects on expected growth rates and invasion speeds differed between examples. The general unstructured model shows no effect, which suggests that demographic structure allows for effects to go in either direction. Positive environmental correlations yielded negative effects on both stochastic growth rates and invasion speeds in the two-stage model, but positive effects in the empirically based \emph{Calathea} model. Similarly, the direction of effects of correlations have been shown to vary in a life history context as well: serial correlations can have strongly positive or negative effects on fitness, depending on the structure of life history and the nature of within-year (among reproductive classes) and between-year correlations in fertilities in response to environmental variation~\citep{tuljapurkar-etal-09}.

We also found that the magnitude of effect of environmental correlations differed between examples. In the \emph{Calathea} example, the effect of correlations was small relative to the importance of the overall proportion of years with positive growth rates, and was more limited than in the two-stage model. Limited effects of temporal environmental correlations on long-term stochastic growth rates have been observed in other empirical studies~\citep{silva-etal-91}, and theoretical studies have found relatively minor effects of temporal correlation in vital rates on stochastic growth rates~\citep{tuljapurkar-82,fieberg-ellner-01}.  However, the effects of temporal autocorrelation can be great under some circumstances, depending on life history and how well vital rates ``remember'' the past~\citep{tuljapurkar-haridas-06}. Populations with weak demographic damping (strong transient effects) tend to accumulate greater variability over time than those with strong damping, and these effects vary with the magnitude of serial correlation~\citep{tuljapurkar-haridas-06}.

Although the \emph{Calathea} example showed relatively small effects of correlations overall, correlations still had the potential to substantially influence growth rates and invasion speed under some conditions, especially for populations near the threshold of positive and negative growth rates. In these threshold cases, strongly positive correlations resulted in the fastest invasion speeds, and also generally the greatest variance in those speeds. This suggests that, on the ground, changes in environmental correlations could lead to sudden or episodic changes in invasion rates for species experiencing threshold growth conditions. These kinds of dynamics might contribute to common observations of a lag between establishment and spread of most introduced species, even those that eventually become invasive, alongside other factors such as adaptation and hybridization with native species~\citep{ellstrand-schierenbeck-00, mack-etal-00,sakai-etal-01}. Among native species responding to climate change, these results suggest that we might expect to observe great variation in rates of range expansion, since populations at the edge of species ranges are most likely to exist close to the threshold between positive and negative population growth. In either case, the increased variability in stochastic growth rates and invasion speed around threshold conditions will make biological expansions and invasions harder to forecast. For species characterized by substantial interannual variation and volatility in vital rates, accurately forecasting rates of spread may be particularly challenging.

Several factors must be taken into careful account in applying integrodifference equation models to invasions on real landscapes. For one, it is critical to understand the basic nature of relevant environmental factors and whether a given form of disturbance imposes different consequences when rare than it does when chronic \citep{tuljapurkar-haridas-06}. For example, Tuljapurkar and Haridas point out that an increase in the frequency of hurricanes may weaken autocorrelation by preventing successional change, thereby effectively reducing demographic variability and the effects of autocorrelation over time despite the outward perception of dramatic environmental variation. Similarly, single years or short-term strings of conditions (e.g., warm years) may boost demographic rates for a given species, while extended autocorrelated series would lead to more substantial ecological changes (e.g., aridification) that negatively impact habitat suitability and represent non-Markovian dynamics. The models we present are applicable to increasing climatic variability but not to sustained directional changes over time. In the context of climate change, these models are likely sufficient to describe many regions of the globe, where increasing climate variability is expected to obscure the effects of concurrent directional trends over much of the next century (IPCC 2007), but not appropriate for regions where trending changes in temperature are more acute.

Despite these limitations, our methods provide a first step in estimating invasion speeds  for populations experiencing stage-specific temporal variation in survivorship, transition rates between stages, fecundity, and dispersal. Using these methods to investigate the effects of temporal correlations on rates of invasion across a wider range of life histories and dispersal types will be an important next step.

\bibliography{../../seb}

\appendix

\section{Mathematical Details}

\renewcommand{\theequation}{\thesection.\arabic{equation}}
We begin by considering solutions to the linear equation 
\begin{equation}\label{eq:linear}
\bn_{t+1}(x)=\int_{-\infty}^\infty \left[ \bK_t(x-y)\circ \bA_t\right] \bn_t (y)\,dy
\end{equation}
Consider $\bn_0(x)= \bu e^{-sx}$ with $\bu>0$. We claim that 
\begin{equation}\label{eq:induct}
\bn_{t}(x)= \bH_{t-1}(s)\dots \bH_0(s) \bu  e^{-sx}
\end{equation}
Indeed, we show this inductively. For $t=1$, 
\begin{eqnarray*}
\bn_{1}(x)&=&\int_{-\infty}^\infty \left[ \bK_0(y-x)\circ \bA_0\right]\bu e^{-sy} \,dy\\
&=& \int_{-\infty}^\infty \left[ \bK_0(z)\circ \bA_0\right] e^{-sz}\,dz\, \bu e^{-sx}\qquad \mbox{where }z=y-x\\
&=&  \bH_0(s) \bu e^{-sx}
\end{eqnarray*}
Next we assume that \eqref{eq:induct} holds for $t-1$. Then 
\begin{eqnarray*}
\bn_{t+1}(x)&=&\int_{-\infty}^\infty \left[ \bK_t(y-x)\circ \bA_t\right]\bn_t(y) \,dy\\
&=&\int_{-\infty}^\infty \left[ \bK_t(y-x)\circ \bA_t\right]\bH_{t-1}(s)\dots \bH_0(s) \bu  e^{-sy} \,dy\\
&=& \int_{-\infty}^\infty \left[ \bK_t(z)\circ \bA_t\right] e^{-sz}\,dz\, \bH_{t-1}(s)\dots \bH_0(s) \bu   e^{-sx}\qquad \mbox{where }z=y-x\\
&=&\bH_t(s)  \bH_{t-1}(s)\dots \bH_0(s)  \bu e^{-sx}
\end{eqnarray*}
which completes the inductive proof of \eqref{eq:induct}.

Let $X_t(s)$ be such that  $\langle\bn_t(X_t(s)),\bw\rangle=n_c$. Then $\langle\bn_0(X_0(s)),\bw\rangle=\langle\bn_t(X_t(s)),\bw\rangle$ and \eqref{eq:induct} imply that
\[
\langle \bu e^{-sX_0(s)} , \bw \rangle = \langle   \bH_{t-1}(s)\dots \bH_0(s) \bu  e^{-sX_t(s)}, \bw \rangle \\
\]
equivalently
\[
X_t(s)-X_0(s) = \frac{1}{s}\ln\frac{\langle   \bH_{t-1}(s)\dots \bH_0(s) \bu, \bw \rangle }{\langle \bu, \bw \rangle}
\]
Equation \eqref{eq:gamma} implies that $\lim_{t\to\infty} \frac{X_t(s)-X_0(s)}{t} = \frac{\gamma(s)}{s} $ with probability one. 

Now consider an initial condition $\bn_0(x)$ with compact support for the non-linear model \eqref{eq:two}. Given any $s>0$, choose $\bu>0$ such that $\bn_0(x)\le \bu e^{-sx}$. Our assumption that $\bB_t(\bn)\le \bA_t$ implies that 
\begin{eqnarray*}
\bn_t(x) \le  \bH_{t-1}(s)\dots \bH_0(s) \bu  e^{-sx}
\end{eqnarray*}
Thus $X_t\le X_t(s)$ where $X_t$ is such that  $\langle\bn_t(X_t),\bw\rangle=n_c$.  Hence, with probability one,
\[
\limsup_{t\to\infty}\frac{X_t-x_0}{t}\le \gamma(s)/s
\]
Since this holds for any $0<s<\hat s$, it follows that 
\[
\limsup_{t\to\infty}\frac{X_t-x_0}{t}\le \min_{0<s<\hat s} \gamma(s)/s
\]
with probability one. The Linearization Conjecture implies that 
\[
\lim_{t\to\infty}\frac{X_t-x_0}{t}= \min_{0<s<\hat s} \gamma(s)/s=:c^*
\]

\section{Log-normal approximation}

Let $s^*$ be such that $\gamma(s^*)/s^*=c^*$. Assume that $\bH_1(s^*), \bH_2(s^*),\dots$ are rapidly mixing (see, e.g., \cite{tuljapurkar-90,heyde-cohen-85} for definitions). Then Theorem 1 in \citep{heyde-cohen-85} implies that  there exists $\sigma\ge 0$ such that 
\[
\frac{\ln \langle  \bH_{t-1}(s^*)\dots \bH_0(s^*) \bu, \bw \rangle -c^*s^* t}{\sqrt{t}\sigma}
\]
converges in distribution to a standard normal as $t\to\infty$.


\end{document}